\documentclass[twocolumn,showpacs,preprintnumbers,amsmath,amssymb]{revtex4}
\usepackage{graphicx}% Include figure files
\usepackage{dcolumn}% Align table columns on decimal point
\usepackage{bm}% bold math

\begin{document}

%\preprint{APS/123-QED}

\title{Knight shift detection using gate-induced decoupling \\of the hyperfine interaction in quantum Hall edge channels}% Force line breaks with \\

\author{S. Masubuchi, K. Hamaya, and T. Machida\footnote{electronic address : tmachida@iis.u-tokyo.ac.jp} }%
\affiliation{Institute of Industrial Science, The University of Tokyo,  \\4-6-1 Komaba, Meguro-ku, Tokyo 153-8505, Japan.
 }%

%%Second institution and/or address\\

\date{\today}% It is always \today, today,

\begin{abstract}
A method for the observation of the Knight shift in nanometer-scale region in semiconductors is developed using resistively detected nuclear magnetic resonance (RDNMR) technique in quantum Hall edge channels. Using a gate-induced decoupling of the hyperfine interaction between electron and nuclear spins, we obtain the RDNMR spectra with or without the electron-nuclear spin coupling. By a comparison of these two spectra, the values of the Knight shift can be given for the nuclear spins polarized dynamically in the region between the relevant edge channels in a single two-dimensional electron system, indicating that this method has a very high sensitivity compared to a conventional NMR technique.
\end{abstract}

                    %display DEGired
\maketitle

%\section{INTRODUCTION}

Confined electrons in two-dimensional semiconductor heterostructures have strong correlation in a high magnetic field, resulting in fascinating spin-related physical phenomena such as spin-texture\cite{Barrett} and quantum Hall (QH) ferromagnetism.\cite{Muraki} Since there is an interplay between electron and nuclear spins through the contact hyperfine interaction in such low-dimensional semiconductors, nuclear magnetic resonance (NMR) spectroscopy is a powerful probe for exploring the strongly interacting electron systems. When the electron spins are polarized around nuclear spins, the nuclear spins are affected by a local effective magnetic field $\left\langle B_\mathrm{e} \right\rangle$ induced by the polarized electron spins, which brings about the change in the Larmor resonance frequency of the nuclear spins in proportional to the electron spin polarization ($P$), i.e. Knight shift.\cite{Slichter} Hence, the observation of the Knight shift has been used to study the skyrmions\cite{Barrett} and the domain structures in QH ferromagnetic states.\cite{Stern}

However, it was difficult for a conventional NMR technique using a pick-up coil to detect NMR signals from the small number of nuclear spins ($N$ $\lesssim$ 10$^{8}$ $\sim$ 10$^{10}$). In this regard, Barrett {\it et al.}\cite{Barrett} used multiple quantum well structures (forty quantum wells) to enhance the NMR signals, and observed the Knight shift in the presence of finite-size skyrmions. On the other hand, Stern {\it et al.}\cite{Stern} recently studied the domain structures in QH ferromagnetic states by means of combining resistively detected NMR (RDNMR) technique \cite{Kronmuller} in a single two-dimensional electron system (2DES) with the conventional NMR technique in a GaAs substrate. For this method, however, it was necessary to compare the RDNMR data with the conventional NMR data to determine the Knight shift.
\begin{figure}[b]
\includegraphics[width=7.5cm]{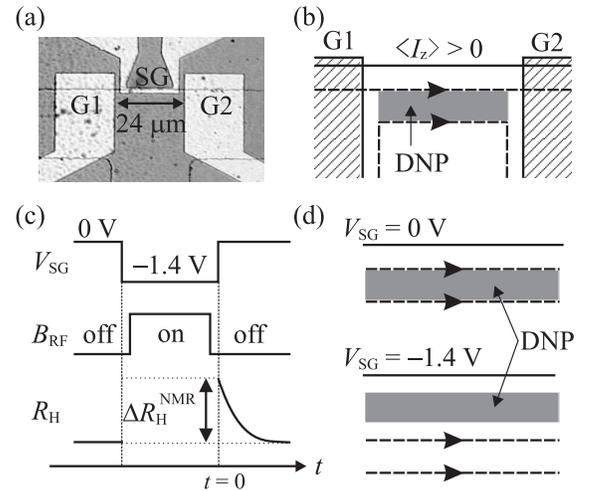}
\caption{(a) An optical micrograph of the sample. (b) Schematic view near the edge region between G1 and G2, where the DNP is induced. (c) Diagram of the experimental procedure for gate-induced decoupling of the hyperfine interaction between electron and nuclear spins. (d) Schematic illustration of the DNP and the edge channels at $V$$_\mathrm{SG}$ = 0 and $-1.4$ V. }
\end{figure}   

In this letter, we develop a method for highly sensitive and resistive detection of the Knight shift in a single two-dimensional electron system ($N$ $\lesssim$ 10$^{8}$) using a gate-induced decoupling of the hyperfine interaction. The NMR spectra observed here are based on only RDNMR technique detected by the change in the Hall resistance ($R_\mathrm{H}$).\cite{Machida1,Machida2} When one applies rf magnetic fields to the polarized nuclear spins, the $R_\mathrm{H}$ increases or decreases at the NMR frequency, resulting from the randomization of the nuclear spin polarization. By comparison of the NMR spectra with and without electrons in the region of detection area, the direct electrical measurements of the Knight shift are possible. This method achieves the detection of the Knight shift for all the kinds of nuclei in GaAs ($^{69}$Ga, $^{71}$Ga, and $^{75}$As) in the nanometer-scale region between QH edge channels in a single 2DES. 

An Al$_{0.3}$Ga$_{0.7}$As/GaAs single-heterostructure crystal with a 2DES ($\mu =$ 40 m$^{2}$/Vs and $n =$ 1.51 $\times$ 10$^{15}$ m$^{-2}$) was fabricated into a Hall-bar device with three Schottky gates (G1, G2, and SG).\cite{Machida2} An optical micrograph of the located gates in the device is shown in Fig. 1(a). The gate-gap length between G1 and G2 is 24 $\mu$m, and a 2-$\mu$m-wide side gate (SG) is located on the edge region of the Hall bar. Transport measurements were performed by standard ac method ($I$$_\mathrm{AC}$ $=$ 1.0 nA) in a $^{3}$He-$^{4}$He dilution refrigerator at 50 mK. In order to induce dynamic nuclear polarization (DNP)  between edge channels, we added $I$$_\mathrm{DC} =$ +7 nA to $I$$_\mathrm{AC}$ (as the electrochemical potential of the outer edge channel with spin-up is larger than that of the inner edge channel with spin-down, we regard the sign of the $I$$_\mathrm{DC}$ as the positive polarity); The positive polarity of $I$$_\mathrm{DC}$ gives rise to the positive DNP ($\left\langle I_\mathrm{z} \right\rangle$ $>$ 0).\cite{Machida1,Machida2} Radio frequency (rf) magnetic fields were applied using a one-turn coil. We hereafter use the condition that the filling factors of Landau levels in the bulk region and under the gates, $\nu_\mathrm{B}$ and $\nu_\mathrm{G}$, are equal to 2 and 1, respectively.
\begin{figure}[t]
\includegraphics[width=8cm]{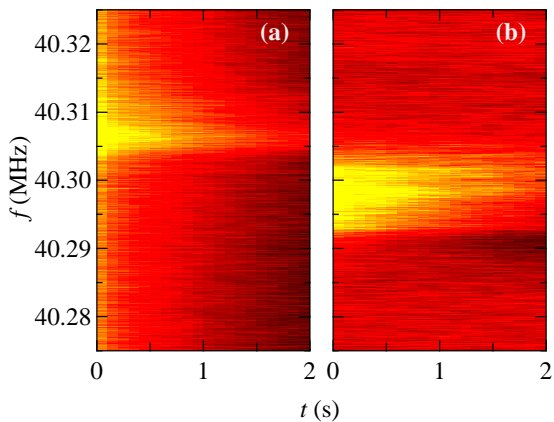}
\caption{Time evolutions of $R_\mathrm{H}$ for various $B_\mathrm{RF}$ frequencies in the cases of (a) $V$$_\mathrm{SG}$ = $-1.4$ V and (b) $V$$_\mathrm{SG}$ = 0 V during the application of the $B_\mathrm{RF}$.}
\end{figure} 

In the quantum Hall state [$B =$ 3.1 T ($\nu_\mathrm{B}$ $=$ 2)], the nuclear spins are polarized dynamically through the hyperfine interaction: when we introduce the difference in the electrochemical potentials between outer and inner edge channels, the spin-flip inter-edge-channel scattering occurs forcibly, giving rise to the DNP between the edge channels.\cite{Machida1,Machida2} The polarized nuclear spins are located in the narrow region ($\sim$ 10 nm) between the spin-resolved edge channels [Fig. 1(b)], where the electron spins are completely polarized ($P$$=$100 \%). To observe the spectrum without electron-nuclear spin coupling, we utilize the depletion of the electron systems around the DNP as follows [Fig. 1(c)]: First of all, the DNP is formed between the edge channels, resulting in a decrease in the $R_\mathrm{H}$ (not shown here).  Secondly, we apply the negative bias voltage ($V$$_\mathrm{SG}$) of $-1.4$ V to the SG. Since the width of the SG (2 $\mu$m) is sufficiently wide with respect to the width of the DNP region ($\sim$ 10 nm), the electron system around the polarized nuclear spins is completely depleted and the relevant edge channels are pushed toward inner region [Fig. 1(d)] at $V$$_\mathrm{SG}$ = $-1.4$ V. Next (0.5 s later), the rf magnetic fields ($B_\mathrm{RF}$) are applied for 2 s. In this procedure, the interplay between electron and nuclear spins is suppressed (gate-induced decoupling). Finally, the $B_\mathrm{RF}$ and $V$$_\mathrm{SG}$ applications are turned off ($t =$ 0), and we begin to record the data of $R_\mathrm{H}$ versus time $t$. Since $I$$_\mathrm{DC}$ $=$ +7 nA remains to be applied, the DNP is formed successively, which leads to a decrease in the $R_\mathrm{H}$. The above measurements are performed continuously with different frequencies $f$. If the $f$ becomes equal to the resonance frequency of each nuclei, the nuclear spins can be depolarized, causing an abrupt increase in the $R_\mathrm{H}$. In order to get the reference data with the electron-nuclear spin coupling, we also measure the $R_\mathrm{H}$ vs $t$ for various $f$ as the $B_\mathrm{RF}$ is applied with $V$$_\mathrm{SG}$ = 0 V.
\begin{figure}[t]
\includegraphics[width=8cm]{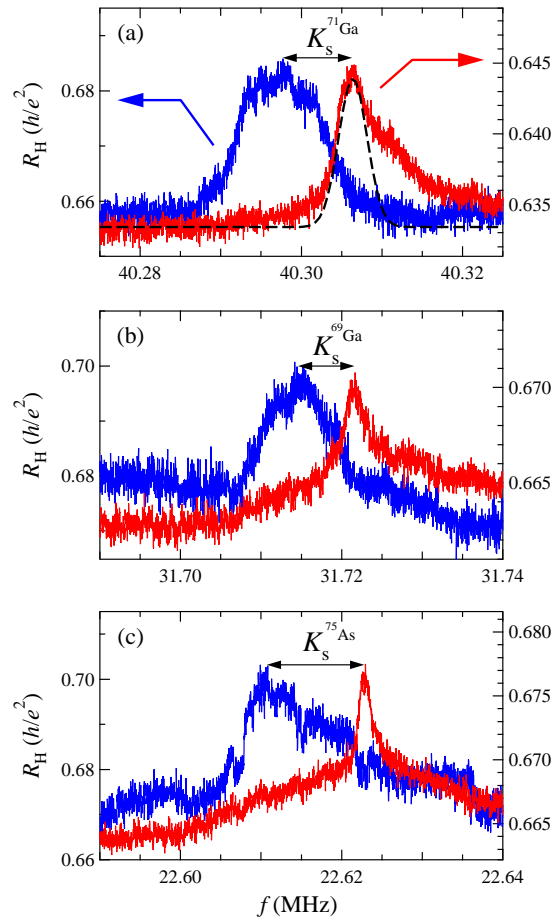}
\caption{NMR spectra at $t =$ 0.5 s with $V$$_\mathrm{SG}$ = $-1.4$ V (red) and $V$$_\mathrm{SG}$ = 0 V (blue) during the $B_\mathrm{RF}$ application for (a) $^{71}$Ga, (b) $^{69}$Ga, and (c) $^{75}$As. The dashed curve (black) is the gaussian curve with the FWHM of 4.2 kHz.}
\end{figure}  

Figures 2(a) and 2(b) show the color plots of $R_\mathrm{H}$ as a function of $t$ for $^{71}$Ga with $B_\mathrm{RF}$ application at (a) $V$$_\mathrm{SG}$ = $-1.4$ V and (b) $V$$_\mathrm{SG}$ = 0 V, respectively. The $R_\mathrm{H}$ values in the yellow regions are larger than those in dark red regions. An yellow region is observed near $f$ $\sim$ 40.305 MHz in Fig. 2(a), while the corresponding yellow region is shifted to $f$ $\sim$ 40.295 MHz in Fig. 2(b). For the other two kinds of nuclei ($^{69}$Ga and $^{75}$As), similar results were also observed.

Figure 3 displays the plots of $R_\mathrm{H}$ vs $f$ at $t =$ 0.5 s in Fig. 2, i.e. the NMR spectra with $B_\mathrm{RF}$ application at $V$$_\mathrm{SG}$ = $-1.4$ V (red  curve) and $V$$_\mathrm{SG}$ = 0 V (blue curve) for (a) $^{71}$Ga, (b) $^{69}$Ga, and (c) $^{75}$As. For all the kinds of nuclei, clear differences in the resonance frequency are seen between the two data. This means that the hyperfine interaction between electron and nuclear spins is turned on or turned off by the side-gate bias voltage, and the difference between these two resonance frequencies corresponds to the value of the Knight shift ($K_\mathrm{S}$), $K_\mathrm{S}$ $=$ (a) 8.6 kHz, (b) 6.9 kHz, and (c) 12.1 kHz. In general, the $K_\mathrm{S}$ is proportional to $P$ and electron density. Considering that the nuclear spins are polarized in the $\nu =$ 1 region between the outer and the inner edge channels, where the $P$ $\sim$ 100 \% and the local electron density is half of that in the bulk region, we infer that these $K_\mathrm{S}$ values are reasonable as compared to the $K_\mathrm{S}$ reported in the previous work at the spin-polarized bulk region of $\nu =$ 1 ($P$ $\sim$ 100 \%) for multiple quantum wells.\cite{Barrett} Also, the relationship between $K_\mathrm{S}$ and the nuclear gyromagnetic constants ($\gamma$$_\mathrm{n}$) is well known as $K_\mathrm{S}$ $=$ $-$$\gamma$$_\mathrm{n}$$\left\langle B_\mathrm{e} \right\rangle$/2$\pi$.\cite{Slichter} In Fig. 3, $K_\mathrm{S}$$^\mathrm{^{71}Ga}$/$K_\mathrm{S}$$^\mathrm{^{69}Ga}$ $\sim$ 1.24 is in good agreement with $\gamma$$_\mathrm{n}$$^\mathrm{^{71}Ga}$/$\gamma$$_\mathrm{n}$$^\mathrm{^{69}Ga}$ $=$ 1.27,\cite{ref} suggesting reliability of this method.
\begin{figure}[t]
\includegraphics[width=8.5cm]{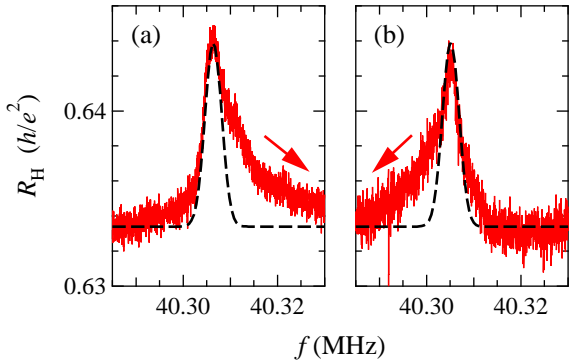}
\caption{NMR spectra with no electron-nuclear spin coupling for $^{71}$Ga in (a) upward and (b) downward sweep of $f$ at 0.2 kHz/min. The curve in (a) corresponds to the red curve shown in Fig. 3(a). The dashed curve (black) is the gaussian curve with the FWHM of 4.2 kHz. }
\end{figure}  

We in turn comment on the change in the NMR line shape and line width (FWHM). Figure 4 shows the sweep direction dependence of the NMR spectra for $^\mathrm{71}$Ga nuclei without electron-nuclear spin coupling at the sweep rate of 0.2 kHz/min. The asymmetry is clearly switched by changing the sweep direction of $f$. This is due to a sufficiently long decay time ($T_\mathrm{1}$) of nuclear spins with respect to the sweep rate of $f$. 
Taking the dependence on the sweep direction into account, we roughly estimate that the FWHM without electron-nuclear spin coupling is $\sim$4.2 kHz, whereas the FWHM in the presence of the electron-nuclear spin coupling is $\sim$11 kHz [blue curve in Fig. 3(a)]. For the other nuclei, the similar features can be seen. These facts reflect that the lower limit of the coherence time ($T_\mathrm{2}$*) is extended owing to the gate-induced decoupling of the hyperfine interaction between electron and nuclear spins. A similar enhancement in the coherence time was recently observed by Yusa {\it et al.}\cite{Yusa} using a fractional quantum Hall device.  

Since the present method is the RDNMR using quantum Hall edge channels in a single 2DES, the small number of the nuclei ($N$ $\lesssim$ 10$^{8}$) is related to the $K_\mathrm{S}$ detection. We deduce that the sensitivity to the nuclei is comparable to the optically detected NMR in a high magnetic field.\cite{Poggio} Also, this study indicates that the gate-bias voltage may systematically control the strength of the hyperfine interaction,\cite{Sanada} suggesting a possibility of the gate-controlled NMR for developing a nuclear spin-based quantum computer.\cite{Kane} Furthermore, a spatial distribution of the electron spin polarization in QH systems can be explored by the gate-controlled hyperfine interaction and the Knight shift detection. 

In summary, we have developed a method for highly sensitive detection of the Knight shift in nanometer-scale region between the QH edge channels for a single two-dimensional electron system using resistively detected NMR. Using gate-induced decoupling of the hyperfine interaction between electron and nuclear spins, we determine the values of the Knight shift for all the kinds of nuclei in GaAs. 

This work is supported by PRESTO, JST Agency, the Grant-in-Aid from MEXT (No. 17244120), and the Support Center for Advanced Telecommunications Technology Research. K. H. acknowledges JSPS Research Fellowships for Young Scientists. 



\begin{thebibliography}{11}
\bibitem{ref}
It is difficult to directly compare $K_\mathrm{S}$$^\mathrm{Ga}$ with $K_\mathrm{S}$$^\mathrm{As}$ because the electron-nuclear coupling constants ($A$) are also related to the $K_\mathrm{S}$.
\end{thebibliography}
\end{document}